\documentclass[
aps,%
12pt,%
final,%
notitlepage,%
oneside,%
onecolumn,%
nobibnotes,%
nofootinbib,%
superscriptaddress,%
centertags]%
{revtex4-1}
\usepackage{amsmath}
\usepackage{amssymb}
\usepackage[dvips]{graphicx}
\begin{document}

\title{THERMAL EFFECTS ON NEUTRINO-NUCLEUS INELASTIC SCATTERING IN STELLAR
ENVIRONMENTS}

\author{\firstname{Alan~A.}~\surname{Dzhioev}}
\email{dzhioev@theor.jinr.ru} \affiliation{\rm Bogoliubov Laboratory
of Theoretical Physics, JINR, 141980, Dubna, Russia.}

\author{\firstname{A.~I.}~\surname{Vdovin}}
\email{vdovin@theor.jinr.ru} \affiliation{\rm Bogoliubov Laboratory of Theoretical Physics, JINR,
141980, Dubna, Russia.}

\author{\firstname{V.~Yu.}~\surname{Ponomarev}}
\email{ponomare@crunch.ikp.physik.tu-darmstadt.de} \affiliation{\rm
Institute for Nuclear Physics, TU Darmstadt, Germany.}

\author{\firstname{J.}~\surname{Wambach}}
\email{Jochen.Wambach@physik.tu-darmstadt.de}\affiliation{\rm
Institute for Nuclear Physics, TU Darmstadt,
Germany.}\affiliation{\rm GSI, Darmstadt, Germany.}

\begin{abstract}
Thermal effects for inelastic neutrino-nucleus scattering off
even-even nuclei in the iron region are studied. Allowed and
first-forbidden contributions to the cross sections are calculated
within the quasiparticle random phase approximation, extended to
finite temperatures within the Thermo-Field-Dynamics formalism. The
GT$_0$ strength distribution at finite temperatures is calculated
for the sample nucleus $^{54}$Fe. The neutral-current
neutrino-nucleus inelastic cross section is calculated for
relevant temperatures during the supernova core collapse. The
thermal population of the excited states significantly enhances the
cross section at low neutrino energies. In agreement with
studies using a large scale shell-model approach the enhancement
is mainly due to neutrino up-scattering at finite temperatures.
\end{abstract}

\maketitle


\section{Introduction}

Neutrinos play a decisive role in core-collapse supernova explosions
since they carry most of the gravitational binding energy released.
The transport of neutrinos through the hot and dense stellar
environment is believed to ultimately be responsible for a
successful explosion, although the details are not fully understood
yet. The present paper addresses  the role of thermal effects in the
inelastic neutrino-nucleus scattering in the iron core during infall
and shortly after bounce.

At the end of 1980th it was pointed out by W.~C.~Haxton that
inelastic neutrino-nucleus scattering (INNS) mediated by the
neutral-current can be of importance comparable with the other
processes of neutrino down-scattering \cite{Haxton88}. The INNS
contributes to the neutrino opacities and thermalization during the
collapse phase, the revival of the stalled shock wave in the delayed
explosion mechanism, and  to explosive nucleosynthesis. The
estimates by Haxton were based on nuclei in their respective ground
states, i.e. for a ``cold'' nuclei. Subsequently, it was realized
that the INNS occurs in hot stellar environment ($T \geq 0.8$~MeV)
and, due to the thermal population of nuclear excited states,
sizeable changes of the INNS cross section are to be expected. The
effect was firstly analyzed in \cite{AstrPhysJ91} and then in
\cite{PLB02} on the basis of large-scale shell-model (LSSM)
calculations. In Refs.~\cite{PLB02,NPA05}, it was found that the
INNS cross section noticeably increases at $T \neq 0$ and for
neutrino energies $E_{\nu} \lesssim 10~\text{MeV}$, especially for
neutrino scattering off even-even nuclides.

However, in the subsequent core-collapse supernova simulations
\cite{PRL08} including several dozens of nuclides, it was
demonstrated that the inclusion of the INNS process does not have
a large effect on the collapse dynamics and the shock wave
propagation. But it significantly modifies the spectrum of
neutrinos generated in the $\nu_e$ burst.

Here, we apply an alternative approach for treating the thermal
effects for INNS cross sections. In essence, our approach is based
on the thermal quasiparticle random phase approximation (TQRPA). We
apply it in the context of Thermo-Field-Dynamics (TFD), which
enables a transparent treatment of thermal excitation and
de-excitation processes and offers the possibility for systematic
improvements. This approach has recently been used in studies of the
electron capture on hot iron and germanium nuclei under stellar
conditions \cite{PRC81}.

\section{FORMALISM}

\subsection{Fundamentals of the Thermo-Field-Dynamics}

Thermo-Field-Dynamics \cite{TFD1,TFD2,Ojima81} is a real-time
formalism for treating thermal effects in quantum field theory and
non-relativistic many-body theories. The standard TFD formalism
treats a many-body
system in thermal equilibrium with a heat bath and a particle
reservoir in the grand canonical ensemble. The thermal average  of a
given operator $A$ is calculated as the expectation value in a
specially constructed, temperature-dependent state $|0(T)\rangle$
which is termed the thermal vacuum. This expectation value is equal
to the usual grand canonical average of $A$. In this sense, the
thermal vacuum describes the hot system in the thermal equilibrium.

To construct the state $|0(T)\rangle$, a formal doubling of the
system degrees of freedom is introduced. In TFD, a tilde conjugate
operator~$\widetilde A$ -- acting in the independent Hilbert space
-- is associated with $A$, in accordance with properly formulated
tilde conjugation rules~\cite{TFD1,TFD2,Ojima81}. For a system
governed  by the Hamiltonian~$H$ at $T=0$, the whole Hilbert space
at $T\neq 0$  is spanned by the direct product of the eigenstates
of~$H$ (${H|n\rangle=E_n|n\rangle}$) and those of the tilde
Hamiltonian~$\widetilde H$ having the same eigenvalues (${\widetilde
H|\widetilde n\rangle=E_n|\widetilde n\rangle}$). The important
point is that, in the doubled Hilbert space, the time-translation
operator is not the initial Hamiltonian~$H$, but instead the thermal
Hamiltonian~${\mathcal H}=H-\widetilde H$. This implies that the
excitations of the thermal system are obtained by the
diagonalization of~${\cal H}$.

The thermal vacuum is the zero-energy eigenstate of the thermal
Hamiltonian $\mathcal H$ and satisfies the thermal state
condition~\cite{TFD1,TFD2,Ojima81}
\begin{equation}\label{TSC}
A|0(T)\rangle = \sigma\,{\rm e}^{{\mathcal H}/2T} {\widetilde
A}^\dag|0(T)\rangle,
\end{equation}
where  $\sigma=1$ for bosonic~$A$ and $\sigma=i$ for fermionic $A$.

As it follows from the definition of $\mathcal H$ each of its
eigenstates with positive energy has the counterpart -- the
tilde-conjugate eigenstate -- with negative but the
same absolute energy value. This allows to treat excitation-
and de-excitation processes at finite temperatures.

Obviously, in most practical cases one cannot diagonalize
$\mathcal H$ exactly. Usually, one resorts to certain approximations
such as Hartree-Fock-Bogoliubov mean field theory (HFB) and the
Random-Phase Approximation (RPA) (see e.g.~\cite{Hat89}). In what
follows the TFD studies for neutrino induced charge-neutral
excitations in hot nuclei are based in part on the results
of~\cite{DzhVdo09,VdoDzh10} (see also~\cite{PRC81}).

\subsection{Charge-neutral excitations in hot nuclei}

In what follows we employ the Hamiltonian of the Quasiparticle-Phonon Model (QPM)
 $H_{\rm QPM}$ \cite{sol92} which consists of proton and neutron mean
fields $H_{\rm sp}$, the BCS pairing interactions $H_{\rm pair}$ and
isoscalar and isovector separable particle-hole interactions. Since
the inelastic neutrino-nucleus scattering  involves nuclear $J^\pi$
excitations of both natural ($\pi=(-1)^J$) and unnatural
($\pi=(-1)^{J+1}$) parities both the separable multipole $H^{\rm
ph}_{\rm M}$ and spin-multipole $H^{\rm ph}_{\rm SM}$ interactions
are included in the particle-hole channel
\begin{equation}\label{QPM}
H_{\rm QPM} = H_{\rm sp} + H_{\rm pair}+ H^{\rm ph}_{\rm M}+H^{\rm
ph}_{\rm SM}.
\end{equation}
The four terms of $H_{\rm QPM}$ read
\begin{align*}
   H_{\rm sp} & = \sum_{\tau=p,n}{\sum_{jm}}^{\tau}(E_{j}-\lambda_\tau)
    a^\dag_{jm}a^{\phantom{\dag}}_{jm}~, \\
H_{\rm pair}& =-\frac14\sum_{\tau=p,n} G_{\tau}{\sum_{\substack{jm \\
j'm'}}}^{\tau}
  a^\dag_{jm}a^\dag_{\overline{\jmath m}}
  a^{\phantom{\dag}}_{\overline{\jmath'm'}}a^{\phantom{\dag}}_{j'm'},
  \\
H^{\rm ph}_{\rm M}&=-\frac12\sum_{\lambda\mu}\sum_{\tau\rho=\pm1}
 (\kappa_0^{(\lambda)}+\rho\kappa_1^{(\lambda)})M^+_{\lambda\mu}(\tau)M^{\phantom{+}}_{\lambda\mu}(\rho\tau)~.
    \\
 H^{\rm ph}_{\rm SM} &=
-\frac12\sum_{L\lambda\mu}\sum_{\tau\rho=\pm1}(\kappa^{(L\lambda)}_0+\rho\kappa^{(L\lambda)}_1)
S^\dag_{L\lambda\mu}(\tau)
S^{\phantom{\dag}}_{L\lambda\mu}(\rho\tau),
\end{align*}
Here, we use standard notations of the QPM. Namely, $a^\dag_{jm}$ and
$a^{\phantom{\dag}}_{jm}$ are the creation and annihilation
operators of particle with quantum numbers $jm\equiv n,l,j,m$ and
energy $E_j$; $\overline{jm}$ stands for the time reversed
single-particle states; the index $\tau$ is isotopic one and
changing the sign of $\tau$ means changing $n \leftrightarrow p$\,;
the parameter $G_\tau$ is the constant of pairing interaction;
$\lambda_\tau$ is the chemical potential; the parameters
$\kappa_{0}^{(a)}$ ($\kappa_{1}^{(a)}$) denote the strength
parameters of the isoscalar (isovector) multipole ($a\equiv\lambda$
is a multipole index)  and spin-multipole ($a\equiv L\lambda$ is a
spin-multipole index) forces. The multipole $M^+_{\lambda\mu}(\tau)$
and spin-multipole $S^+_{L\lambda\mu}(\tau)$ single-particle
operators read as
 \begin{align}\label{mult}
 M^+_{\lambda\mu}(\tau)&={\sum_{\genfrac{}{}{0pt}{1}{j_1m_1}{j_2m_2}}}^{\tau}
 \langle j_1m_1|i^\lambda R_\lambda(r)Y_{\lambda\mu}|j_2m_2\rangle
 a^\dag_{j_1m_1}a^{\phantom{\dag}}_{j_2m_2}~,
 \notag   \\
 S^\dag_{L\lambda\mu}(\tau) &= {\sum_{\genfrac{}{}{0pt}{1}{j_1m_1}{j_2m_2}}}^{\tau}
 \langle j_1m_1| i^L R_L(r) [Y_{L}\vec\sigma]^\lambda_\mu|j_2m_2\rangle
  a^\dag_{j_1m_1}a^{\phantom{\dag}}_{j_2m_2}\, ,
  \end{align}
where
\begin{equation*}
  \bigl[Y_L\,\sigma\bigr]^\lambda_\mu=\sum_{M,\,m}\langle LM\,1m|\lambda\mu\rangle
 Y_{LM}(\theta,\phi)\sigma_m~,
\end{equation*}
and the notation ${\sum}^\tau$  implies a summation over neutron
($\tau=n$) or proton ($\tau=p$) single-particle states only. The
excitations of natural parity are generated by the multipole and
spin-multipole $L=\lambda$ interactions, while the spin-multipole
interactions with $L=\lambda\pm1$ are responsible for the states of
unnatural parity.

To determine the thermal behavior of a nucleus governed by the
Hamiltonian (\ref{QPM}) we should diagonalize  the thermal
Hamiltonian $\mathcal{H}_{\rm QPM} = H_{\rm QPM} -
\widetilde{H}_{\rm QPM}$ and find the corresponding thermal vacuum
state. This will be done in two steps.

In a  first step, the sum  of single-particle and pairing
 terms $\mathcal{H}_{\rm BCS}=\mathcal{H}_{\rm sp} +\mathcal{H}_{\rm pair}$
is diagonalized.  To this end two subsequent unitary transformations
are made. The first is the usual Bogoliubov $u, v$ transformation
from the original particle operators
$a^\dag_{jm},~a^{\phantom{\dag}}_{jm}$ to the quasiparticle ones
$\alpha^\dag_{jm},~\alpha^{\phantom{\dag}}_{jm}$. The same
transformation is applied to the tilde operators $\widetilde
a^\dag_{jm},\ \widetilde a^{\phantom\dag}_{jm}$, thus producing the
tilde quasiparticle operators $\widetilde\alpha^\dag_{jm},\
\widetilde\alpha^{\phantom\dag}_{jm}$. The second, unitary thermal
Bogoliubov transformation mixes the original and tilde degrees of
freedom
\begin{align}\label{TBt}
       \beta^\dag_{jm}&=x_j\alpha^\dag_{jm}\!-\!i y_j\widetilde\alpha_{jm}\\
       \widetilde\beta^\dag_{jm}&=x_j\widetilde\alpha^\dag_{jm}\!+\!i
       y_j\alpha_{jm}~~~(x^2_j+y^2=1). \nonumber
\end{align}
The operators $\beta^\dag_{jm}, \beta_{jm}, \widetilde
\beta^\dag_{jm}$,  and $\widetilde\beta_{jm}$ are called thermal
quasiparticle operators.

The coefficients $u_j,\ v_j,\ x_j,\ y_j$  are found by
diagonalizing~${\cal H}_{\rm BCS}$ and demanding that the vacuum of
thermal quasiparticles is the thermal vacuum in the  BCS
approximation, i.e., it obeys the thermal state
condition~\eqref{TSC}.  As a result one obtains the following
equations for $u_j,\ v_j$ and $x_j,\ y_j$:
\begin{eqnarray}
v_j & = & \frac{1}{\sqrt
2}\left(1-\frac{E_j-\lambda_{\tau}}{\varepsilon_j}\right)^{1/2},\
 u_j=(1-v_j^2)^{1/2}, \label{u&v} \\
y_j & = &
\left[1+\exp\left(\frac{\varepsilon_j}{T}\right)\right]^{-1/2},\
  x_j=\bigl(1-y^2_j\bigr)^{1/2}, \label{x&y}
\end{eqnarray}
where $\varepsilon_j=\sqrt{(E_j-\lambda_{\tau})^2+\Delta^2_{\tau}}$.
The coefficients $y^2_j$ determine the average number of thermally
excited Bogoliubov quasiparticles in the BCS thermal vacuum
\begin{equation}
  \langle 0(T);{\rm qp}|
  \alpha^\dag_{jm}\alpha^{\phantom{\dag}}_{jm}
  |0(T);{\rm qp}\rangle=y^2_{j}
\end{equation}
 and, thus, coincide with the thermal occupation factors of the
Fermi-Dirac statistics.

The pairing gap $\Delta_{\tau}$  and the chemical potential
$\lambda_\tau$ are the solutions to the finite-temperature BCS
equations
\begin{align}\label{BCS}
\Delta_\tau(T)&=\frac{G_\tau}{2}{\sum_j}^\tau(2j+1)(1-2y^2_j)u_jv_j,\nonumber\\
N_\tau&={\sum_j}^\tau(2j+1)(v^2_jx^2_j+u^2_jy^2_j),
\end{align}
where $N_\tau$ is the number of neutrons or protons in a nucleus.

At this stage, the thermal BCS Hamiltonian ${\mathcal H}_{\rm BCS}$
is diagonal
\begin{equation*}
{\mathcal H}_{\rm BCS}
\simeq\sum_\tau{\sum_{jm}}^\tau\varepsilon_j(T)
(\beta^\dag_{jm}\beta^{\phantom{\dag}}_{jm}-\widetilde\beta^\dag_{jm}\widetilde\beta^{\phantom{\dag}}_{jm}),
\end{equation*}
and corresponds to a system of non-interacting thermal
quasiparticles.  The vacuum for thermal quasiparticles ${|0(T);{\rm
qp}\rangle}$ is the thermal vacuum in the BCS approximation. The
states $\beta^\dag_{jm}|0(T);{\rm qp}\rangle$ have positive
excitation energies whereas the corresponding tilde-states
$\widetilde\beta^\dag_{jm}|0(T);{\rm qp}\rangle$ have negative
energies. Since the thermal vacuum contains a certain number of
Bogoliubov quasiparticles, excited states can be built on
$|0(T);{\rm qp}\rangle$ by either adding or removing a Bogoliubov
quasiparticle. The first process corresponds to the creation of a
non-tilde thermal quasiparticle with positive energy, whereas the
second process creates a tilde quasiparticle with negative energy.

At the second step of the approximate diagonalization of ${\mathcal
H}_{\rm QPM}$, long-range correlations due to the particle-hole
interaction are taken into account within the thermal QRPA (TQRPA).
Within the TFD formalism the terms ${\cal H}^{\rm ph}_{\rm M}$ and
${\cal H}^{\rm ph}_{\rm SM}$  are written in terms of the thermal
quasiparticle operators determined above. Then, ${\cal H}_{\rm QPM}$
is approximately diagonalized within a basis of thermal phonon
operators
 \begin{multline}\label{phonon}
  Q^\dag_{\lambda \mu i}=\frac12\sum_\tau{\sum_{j_1j_2}}^\tau
 \Bigl\{\psi^{\lambda i}_{j_1j_2}[\beta^\dag_{j_1}\beta^\dag_{j_2}]^\lambda_\mu +
 \widetilde\psi^{\lambda i}_{j_1j_2}[\widetilde\beta^\dag_{\overline{\jmath_1}}
 \widetilde\beta^\dag_{\overline{\jmath_2}}]^\lambda_\mu +
 2i\,\eta^{\lambda i}_{j_1j_2}[\beta^\dag_{j_1}
  \widetilde\beta^\dag_{\overline{\jmath_2}}]^\lambda_\mu\\
+
 \phi^{\lambda i}_{j_1j_2}[\beta_{\overline{\jmath_1}}\beta_{\overline{\jmath_2}}]^\lambda_{\mu} +
 \widetilde\phi^{\lambda i}_{j_1j_2}[\widetilde\beta_{j_1}
 \widetilde\beta_{j_2}]^\lambda_{\mu} +
 2i\,\xi^{\lambda i}_{j_1j_2}[\beta_{\overline{\jmath_1}}
  \widetilde\beta_{j_2}]^\lambda_{\mu}\Bigr\},
\end{multline}
where $[~]^\lambda_\mu$ denotes the coupling of single-particle
angular momenta $j_1, j_2$ to a total angular momentum $\lambda$.
Now the thermal equilibrium state is treated as the vacuum
$|0(T);{\rm ph}\rangle$ for the thermal phonon annihilation
operators.

The thermal phonon operators are considered as bosonic ones which
imposes certain constraint on the phonon amplitudes. To find the
amplitudes and energies of the thermal phonons, the variational
principle is used, i.e., we  find the minimum of the average value
of thermal Hamiltonian with respect to the one-phonon
states~${Q^\dag_{\lambda\mu i}|0(T);{\rm ph}\rangle}$ or
${\widetilde Q^\dag_{\overline{\lambda\mu i}}|0(T);{\rm ph}\rangle}$
under the aforementioned constraint.

After variation one obtains a system of linear equations for the
amplitudes $\psi^{\lambda i}_{j_1j_2},\ \widetilde\psi^{\lambda
i}_{j_1j_2},\ \eta^{\lambda i}_{j_1j_2}$, etc. as well as for the energies
(details can be found in ref. \cite{DzhVdo09}).
These constitute the equations for the thermal quasiparticle
random phase approximation. In contrast to the zero temperature
case, the negative solutions of the secular equation
have a physical meaning. They correspond to the tilde thermal
one-phonon states and arise from
$\widetilde\beta^\dag\widetilde\beta^\dag$ terms in the thermal
phonon operator. As it was noted above, creation of a tilde thermal
quasiparticle corresponds to the annihilation of a thermally excited
Bogoliubov quasiparticle. Consequently, excitations of
negative-energy thermal phonons correspond to transitions from
thermally excited nuclear states.

After diagonalization in terms of thermal phonon operators the TQRPA
part of the ${\cal H}_{\rm QPM}$ takes the form
\begin{equation}
{\cal H}_{\rm TRPA}=\sum_{\lambda\mu i}\omega_{\lambda i}
   (Q^\dag_{\lambda\mu i}Q^{\phantom{\dag}}_{\lambda\mu i}
   -\widetilde Q^\dag_{\lambda\mu i}\widetilde Q^{\phantom{\dag}}_{\lambda\mu i}).
\end{equation}
To fix properly the thermal vacuum state $|0(T);{\rm ph}\rangle$
corresponding to TRPA we once again turn to the thermal state
condition \eqref{TSC} and derive the final expressions for the
amplitudes of the thermal phonon operator \eqref{phonon}.

Once the structure of thermal phonons is determined, one can determine
the transition probabilities from the thermal vacuum to thermal
one-phonon states. They are given by the squared reduced matrix
elements of the corresponding transition operator
$\mathcal{T}_{\lambda\mu}$
\begin{align}\label{trans_ampl}
\Phi_{\lambda i}&=\bigl|\langle Q_{\lambda
i}\|\mathcal{T}_{\lambda}\|0(T);\mathrm{ph}\rangle\bigr|^2,
  \notag\\
\widetilde \Phi_{\lambda i}&=\bigl|\langle\widetilde Q_{\lambda
i}\|\mathcal{T}_{\lambda}\|0(T);\mathrm{ph}\rangle\bigr|^2.
\end{align}
Thus,  the probability to excite the hot nucleus is given by
$\Phi_{\lambda i}$,  while $\widetilde\Phi_{\lambda i}$  is the
probability to de-excite it.

\subsection{Cross section of inelastic neutrino-nucleus scattering}

Considering neutrino-nucleus inelastic scattering in stellar
environments we assume that a nucleus is in thermal equilibrium
with a heat bath and particle reservoir
or, in TFD terms, in the thermal (phonon) vacuum state. An
inelastic collision of a hot nucleus with neutrinos leads to
transitions from the thermal vacuum to thermal one-phonon states.

In the derivation of the relevant cross section at finite
temperature  we follow the formalism by Walecka-Donnelly \cite{Wal75,Don79},
which describes in a unified way electromagnetic and weak
semileptonic processes by taking advantage of the
multipole decomposition of the relevant hadronic current density
operator. In the case of  neutral-current neutrino-nucleus
scattering, the differential cross section for  a transition from an
initial nuclear state ($i$) to a final state ($f$) can be written as
a sum over all allowed multipolarities $J^\pi$

\begin{equation}\label{dif_cr_sect}
  \frac{ d\sigma_{i\to f}}{d\Omega} = \frac{2G^2}{\pi}\frac {(E_\nu-\omega_{if})^2 \cos^2\frac{\Theta}{2} }{2J_i+1}
    \Bigl\{\sum_{J=0}^\infty \sigma^J_{\rm CL} + \sum_{J=1}^\infty \sigma^J_{\rm T}
    \Bigr\},
\end{equation}
where
\begin{equation}\label{CL}
  \sigma^J_{\rm CL} = \big|\langle J_f\| \hat M_J + \frac{\omega_{if}}{q} \hat L_J\|J_i\rangle \big|^2
\end{equation}
and
\begin{multline}\label{T}
 \sigma^J_{\rm T}=\Bigl(-\frac{q^2_\mu}{2q^2} + \tan^2\frac{\Theta}{2} \Bigr)
  \Bigl[ |\langle J_f\| \hat J^{\rm mag}_J\|J_i\rangle|^2 + |\langle J_f\| \hat J^{\rm el}_J\|J_i\rangle|^2       \Bigr]
     \\
   -\tan\frac{\Theta}{2}\sqrt{-\frac{q^2_\mu}{2q^2} + \tan^2\frac{\Theta}{2}}
    \Bigl[ 2 \mathrm{Re} \langle J_f\| \hat J^{\rm mag}_J\|J_i\rangle\langle J_f\| \hat J^{\rm el}_J\|J_i\rangle^*  \Bigr].
\end{multline}
Here $G$ is the electroweak coupling constant, $\Theta$ is the
scattering angle, $E_\nu$ is the incoming neutrino energy,
$\omega_{if}$ is the transition energy from the initial nuclear
state ($i$) to the final state ($f$), and $q_\mu=(\omega_{if}, \vec
q)$ $\Big(q=|\vec
q|=\sqrt{\omega_{if}^2+4E_\nu(E_\nu-\omega_{if})\sin^2\frac{\Theta}{2}}~\Big)$
is the four-momentum transfer. The operators $\hat M_J$, $\hat L_J$,
$\hat J^{\rm el}_J$, and $\hat J^{\rm mag}_J$ are the multipole
operators for the charge, longitudinal,  and the transverse electric
and magnetic parts of the four-current, respectively. Following
\cite{Wal75} they can be written in terms of one-body operators in
the nuclear many-body Hilbert space.

The cross section involves the reduced matrix elements of these
operators between the initial and final nuclear states. Within the
present approach, the initial nuclear state is the thermal phonon
vacuum (TV) and the final states are the thermal one-phonon states.
Therefore, at $T\neq 0$ all the reduced matrix elements in
Eqs.~(\ref{CL},\ref{T}) are calculated in accordance with
Eqs.~\eqref{trans_ampl}. The total cross section is obtained from
the differential  cross sections by summing over all possible
one-phonon states of different multipolarity  and by numerical
integration over scattering angles
\begin{equation}
  \sigma(E_\nu)=2\pi\sum_{f\in\{\lambda i\}} \int_1^{-1} \frac{ d\sigma_{\mathrm{TV}\to f}}{d\Omega}d \cos\Theta.
\end{equation}

Up to moderate  energies ($E_{\nu}\sim 15-20$~MeV), the inelastic
neutrino-nucleus scattering is dominated by the neutral-channel
Gamow-Teller transitions $J^\pi=1^+$. Moreover, in the $q \to 0$ limit,
the full operator exciting $1^+$ states is reduced to the following
Gamow-Teller operator:
\begin{equation}\label{GT0}
\mathrm{GT}_0= \Big(\frac{g_{A}}{g_{V}}\Bigr)\vec\sigma t_0
 \end{equation}
where $(g_{A}/g_{V})=-1.2599$~\cite{Towner95} is the ratio of the
axial and vector weak coupling constants, $\vec \sigma$ is the spin
operator and $t_0$ is the zero-component of the isospin operator in
spherical coordinates.

To circumvent computational limitations in the LSSM calculations
\cite{PLB02,NPA05,PRL08} the total INNS  cross section
$\sigma(E_\nu)$ was split into two parts -- a down-scattering part
$\sigma_{d}(E_\nu)$ and the up-scattering part $\sigma_{u}(E_\nu)$.
The term $\sigma_d(E_\nu)$ includes transitions where the scattered
neutrino loses energy whereas the term $\sigma_{u}(E_\nu)$ includes
those transitions where the neutrino gains energy from a hot
nucleus. Assuming the validity of the Brink hypothesis for the
GT$_0$ resonance, the down-scattering term was transformed to a sum
over only those final excited nuclear states which are coupled by a
direct GT$_0$ transition with the nuclear ground state. As a result,
$\sigma_{d}(E_\nu)$ appeared to be independent of $T$.

In our case, the part $\sigma_d(E_\nu)$ corresponds to transitions
from $|0(T);\mathrm{ph}\rangle$ to $|Q_{\lambda i}\rangle$ states
with positive energies whereas the $\sigma_{u}(E_\nu)$ term is the
sum of transitions $|0(T);\mathrm{ph}\rangle \to
|\widetilde{Q_{\lambda i}}\rangle$ where the tilde-states have negative
energies. In the latter transitions a neutrino gains energy due
to nuclear de-excitation.

Thus within the present approach the GT$_0$ ($J^\pi=1^+$)
contribution to the cross section reads
\begin{equation}\label{INNS-TFD}
\sigma(E_\nu)=\sigma_d(E_\nu)+\sigma_{u}(E_\nu) =
\frac{G^2}{\pi}\sum_{i}(E_\nu -\omega_{J i})^2\Phi_{J
i}+\frac{G^2}{\pi}\sum_{i}(E_\nu +\omega_{J i})^2\widetilde\Phi_{J
i},
\end{equation}
The probabilities $\Phi_{J i}$ and $\widetilde\Phi_{J i}$ are given
in \eqref{trans_ampl} with $\mathcal{T} = \mathrm{GT_0}$. Since
$\omega_{J i}$,  $\Phi_{J i}$ and $\widetilde\Phi_{J i}$ are
functions of $T$, both terms $\sigma_d$ and $\sigma_u$ depend on
temperature.

Whereas the GT$_0$ component determines the neutrino-nucleus cross
section at low $E_\nu$,  higher multipole contributions become
increasingly important at higher neutrino energies. Moreover, at
higher neutrino energies Eq.~\eqref{GT0} for GT$_0$ is not valid and
the $1^+$ transition operator will depend on transfer momentum $q$.
According to Refs.~\cite{Kolbe1999, Hektor2000} the $q$-dependence
reduces the cross section.

\section{Calculations for the hot nucleus $^{54}$Fe}

Numerical calculations have been performed for $^{54}$Fe.
The single-particle wave functions and energies were calculated in a
spherically symmetric Woods-Saxon potential. The constants of the
pairing interaction were determined to reproduce experimental
pairing energies in the BCS approximation. All parameters are
the same as in our previous calculations \cite{PRC81,BulRAS08} for
electron capture rates on the same nucleus at $T\neq 0$.

The radial dependence of the residual multipole and spin-multipole
forces is chosen in the form $R_\lambda(r)=\partial U(r)/\partial r$
where $U(r)$ is the central part  of the single-particle Woods-Saxon
potential. Thus, $R_\lambda(r)$ as well as the parameters
$\kappa^{(\lambda)}_{0,1}$ and $\kappa^{(L\lambda)}_{0,1}$ do not
depend on $\lambda$. The isovector parameters
$\kappa^{(\lambda)}_{1}$ and $\kappa^{(L\lambda)}_{1}$ are fitted to
the experimental position of the E1~\cite{Nor78} and
M1~\cite{Richter85} resonances in $^{54}$Fe. According to the
estimates in Refs.~\cite{DTKPV86,PVV87}, the isoscalar
spin-multipole interaction is very weak in comparison with the
isovector one. Following~\cite{PVV87}, we take
$\kappa^{(L\lambda)}_{0}/\kappa^{(L\lambda)}_{1}=0.1$.

\begin{figure*}[h]
\includegraphics[width=\textwidth]{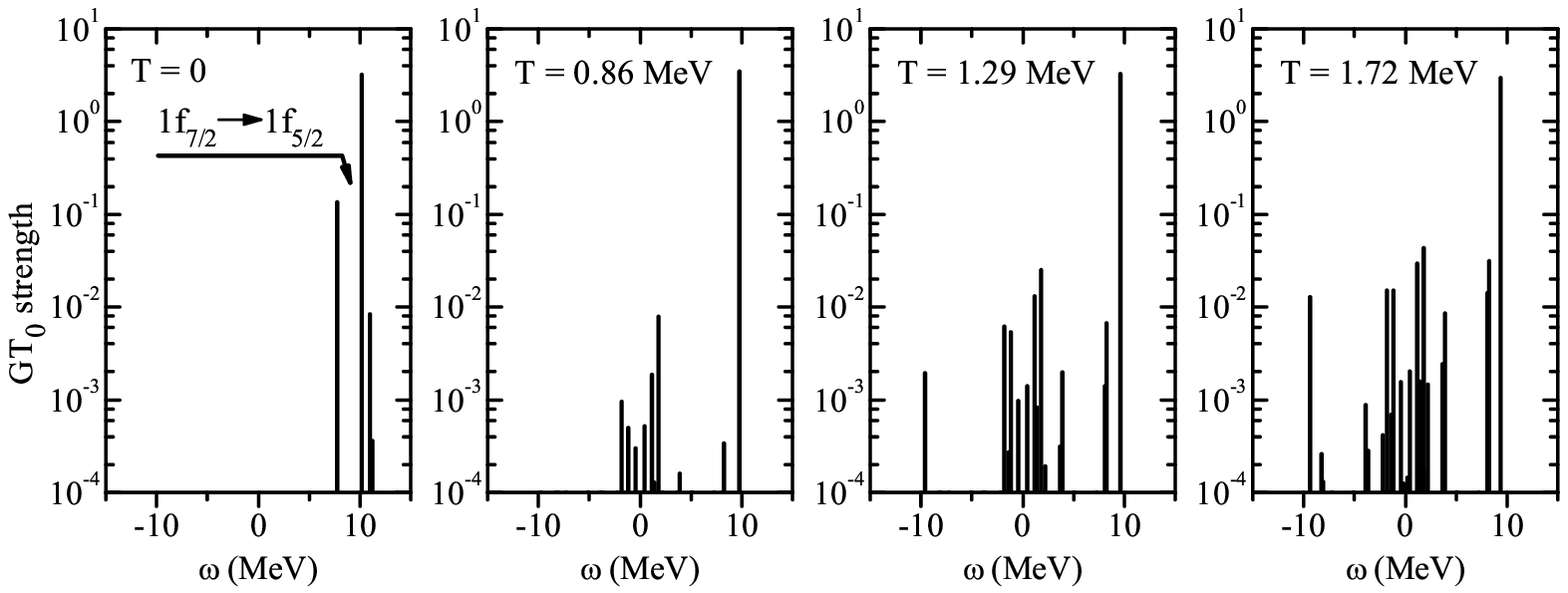}
\caption{}\label{figure1}
\end{figure*}

First, we have performed TQRPA calculations of the GT$_0$ strength
distribution in $^{54}$Fe.   As in the LSSM
calculations~\cite{NPA05}, the GT$_0$ operator~\eqref{GT0} have been
scaled by a quenching factor~$0.74$. In Fig.~\ref{figure1}, we
display the GT$_0$ strength distributions for the ground state
($T=0$)  of $^{54}$Fe and at three stellar temperature values,
occurring at different collapse stages: $T=0.86$~MeV corresponds to
the condition in the core of a presupernova model for a
$15\text{M}_{\odot}$ star; $T=1.29$~MeV and $T=1.72$~MeV relate
approximately  to the neutrino trapping and neutrino thermalization
stages, respectively. All results are plotted as a function of the
energy transfer to $^{54}$Fe. For charge-neutral reactions this
energy is  equal to a thermal phonon energy $\omega_{J i}$.

At $T=0$, the transition strength is concentrated mostly in
one-phonon $1^+$ state forming the GT$_0$ resonance near $\omega
\approx 10$~MeV. The main contribution to the phonon structure comes
from the proton and neutron single-particle transitions $1f_{7/2}
\to 1f_{5/2}$.  With temperature increase the fraction of low-energy
transitions in the GT$_0$ strength distribution increases. The physical
reason is the weakening and subsequent collapse of pairing
correlations (at $T \approx 0.8$~MeV) and appearance of low-energy
particle-particle and hole-hole transitions due to thermal smearing
of neutron and proton Fermi surfaces. Moreover, at finite
temperature the ``negative energy'' transitions to tilde one-phonon
states appear.  As a result, the GT$_0$ energy centroid is shifted
down by 1.1~MeV at $T=1.72$~MeV. This indicates a violation of the
Brink hypothesis within the present approach.

\begin{figure*}[h]
\includegraphics[width=\textwidth]{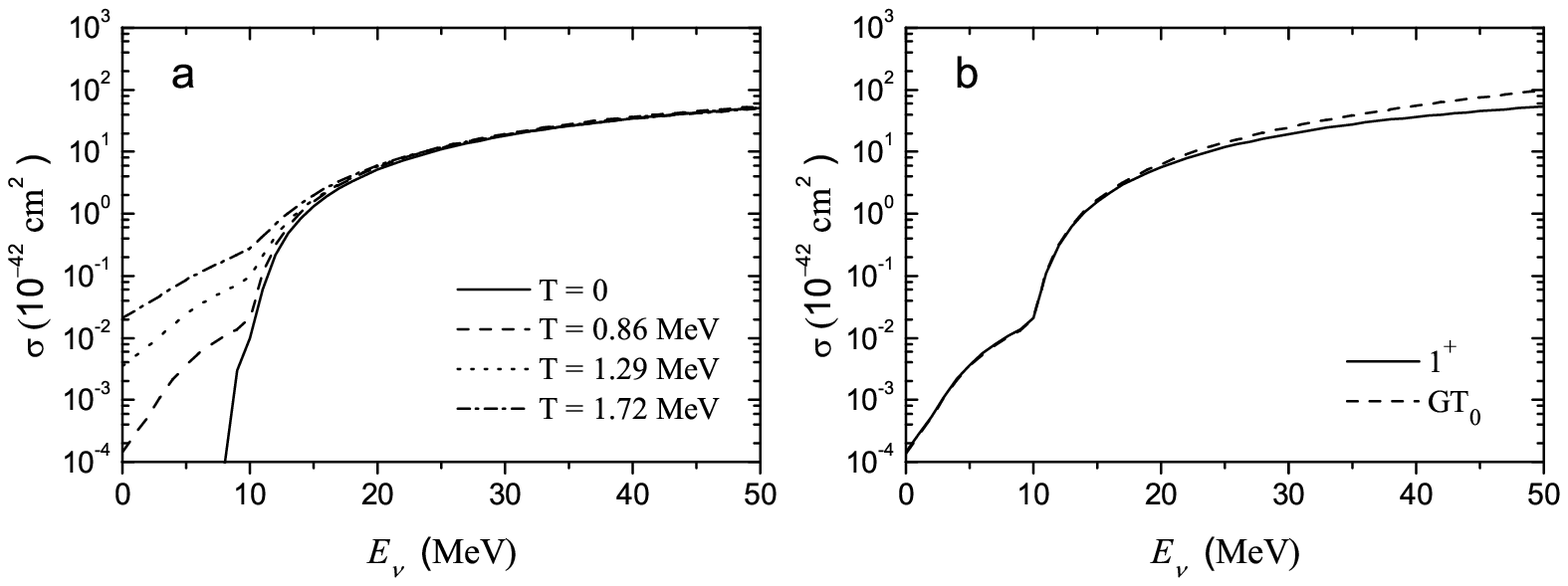}
\caption{}\label{figure2}
\end{figure*}

The contribution of $1^+$ transitions to  the INNS cross section is
shown in Fig.~\ref{figure2}(a) for different temperatures. The
calculations have been performed with the exact $q$-dependent $1^+$
multipole transition operator~\cite{Wal75}. As in the LSSM
calculations \cite{PLB02}, the cross section $\sigma(E_\nu)$ at
$T=0$ is equal to zero when $E_\nu$ is less than the energy of the
lowest $1^+$ state in $^{54}$Fe. Within the QRPA, the lowest $1^+$
state in $^{54}$Fe has an excitation energy of $\omega(1^+)\approx
7.5$~MeV (see Fig.~\ref{figure1}). The GT$_0$ transitions at $T\ne
0$ do not show such a gap due to thermally unblocked low- and
negative-energy transitions.  As a consequence, there is no a
threshold energy for neutrinos at finite temperatures and the INNS
cross section appears to be quite sensitive to $T$ at neutrino
energies $E_\nu<10$~MeV. As it follows from the present calculations
as well as from the LSSM study~\cite{PLB02}, thermal effects can
increase the low energy cross section by up to two orders of
magnitude when the temperature rises from 0.86~MeV to 1.72~MeV.
Finite temperature effects are unimportant for $E_\nu>15$~MeV where
excitation of the GT$_0$ resonance becomes possible and dominates
the cross section. These features were pointed in \cite{PLB02} as
well.

To check the influence of finite momentum transfer on the INNS cross
section we also have performed calculations with the GT$_0$
transition operator~\eqref{GT0}. A comparison of $1^+$ and GT$_0$
cross sections is shown in Fig.~\ref{figure2}(b) for $T=0.86$~MeV.
The $q$-dependence becomes important at $E_\nu>30$~MeV. At
$E_\nu=35$~MeV the INNS cross section calculated with the
$q$-dependent $1^+$ operator is by 20\% less than that calculated
with the GT$_0$ operator \eqref{GT0}. At $E_\nu=50$~MeV the
difference is by about factor of 2. The effect does not change with
temperature.

\begin{figure*}[h]
\includegraphics[width=\textwidth]{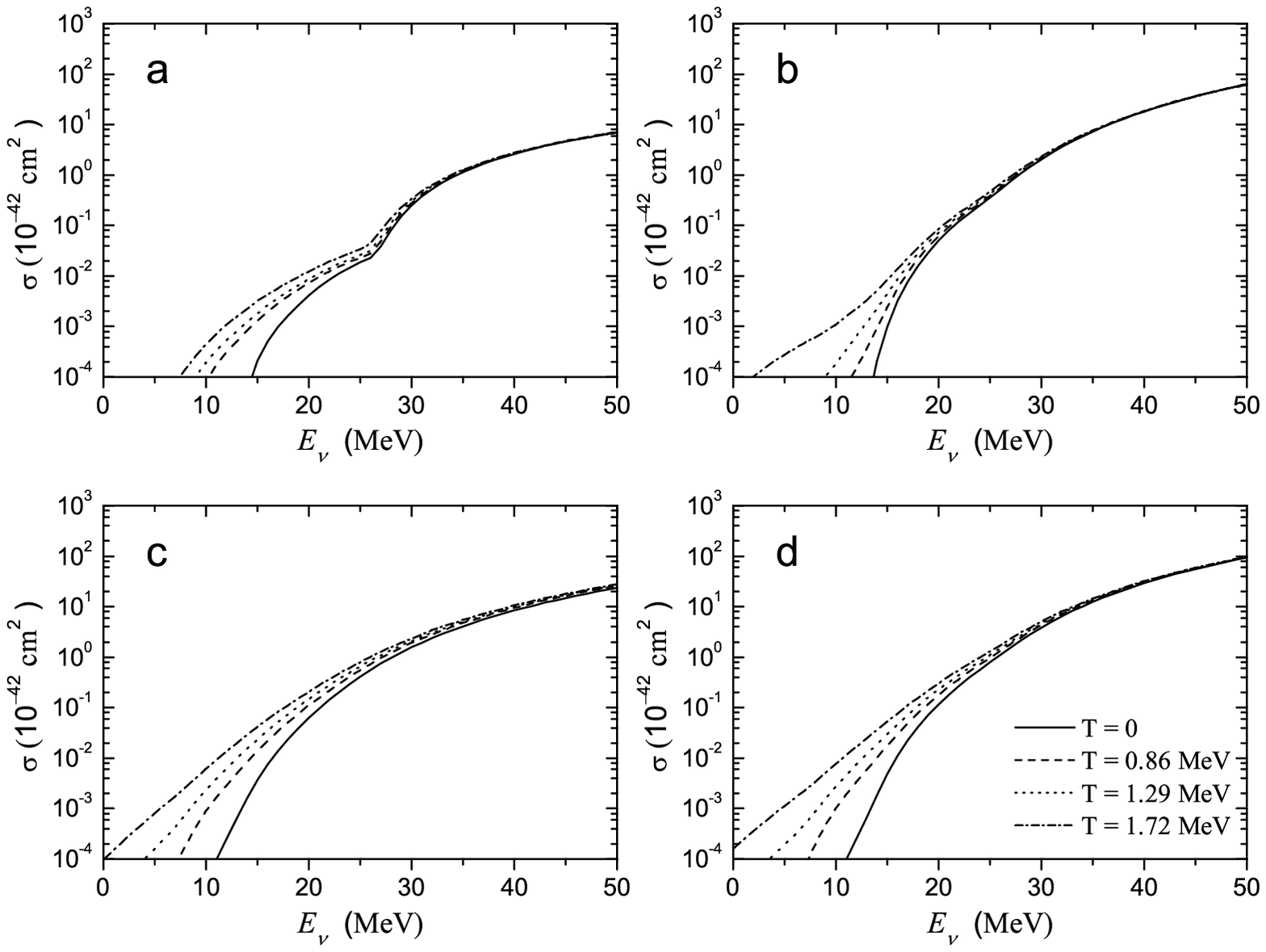}
\caption{}\label{figure3}
\end{figure*}

The contribution of first-forbidden transitions $0^-,~1^-$, and
$2^-$  to the INNS cross section were also calculated within the TQRPA,
taking into account the $q$-dependence as given in~\cite{Wal75}. The
results are presented in Fig.~\ref{figure3}. As it can be seen, a
temperature increase enhances the cross sections at low
and moderate $E_\nu$. The main reason is thermally unblocked
low-energy first-forbidden transitions. According to our
calculations  $2^-$ transitions dominate the total contribution of
first-forbidden transitions to the cross section at low neutrino
energies, while at higher energies the total contribution is
mainly determined by the $1^-$ transitions.

\begin{figure*}[h]
\includegraphics[width=0.6\textwidth]{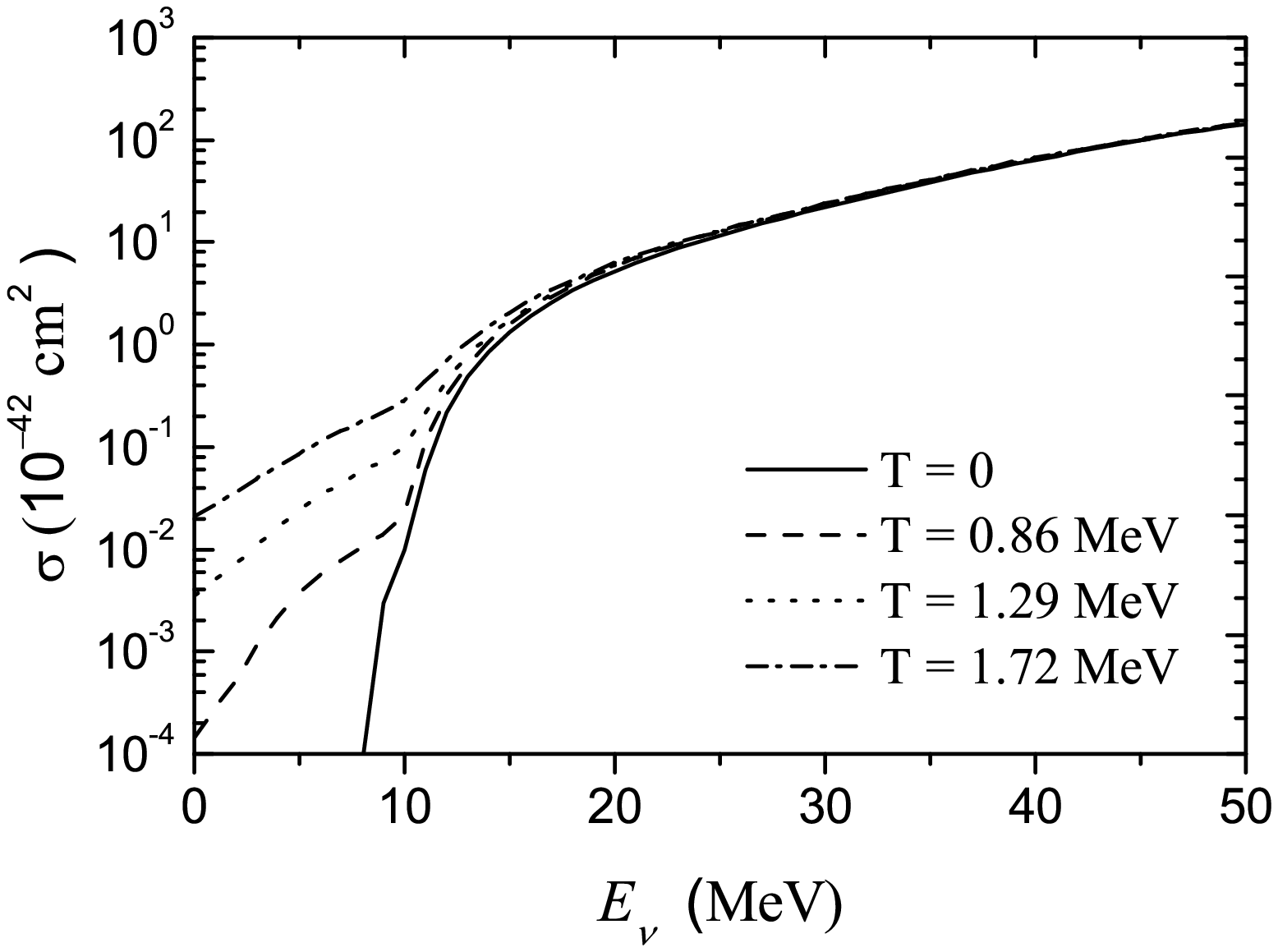}
\caption{}\label{figure4}
\end{figure*}

In Fig.~\ref{figure4}, the INNS cross sections at different
temperatures are shown as a sum of $1^+, 0^-,1^-$, and $2^-$
contributions (we omit the contribution of the $0^+$ multipole
because it is negligible). At low $E_\nu$ the cross sections are
almost completely dominated by the GT$_0$ transitions. The part of
the cross sections arising from the first-forbidden transitions
becomes increasingly important at larger $E_\nu$. We find that for
$E_\nu=30$~MeV up to 20\% of the cross section is due to
first-forbidden transitions. For $E_\nu=40$~MeV allowed and
forbidden transitions contribute about equally, while at
$E_\nu=50$~MeV the contribution of first-forbidden transitions is
nearly twice as large as that of $1^+$ transitions.

\begin{figure*}[h]
\includegraphics[width=\textwidth]{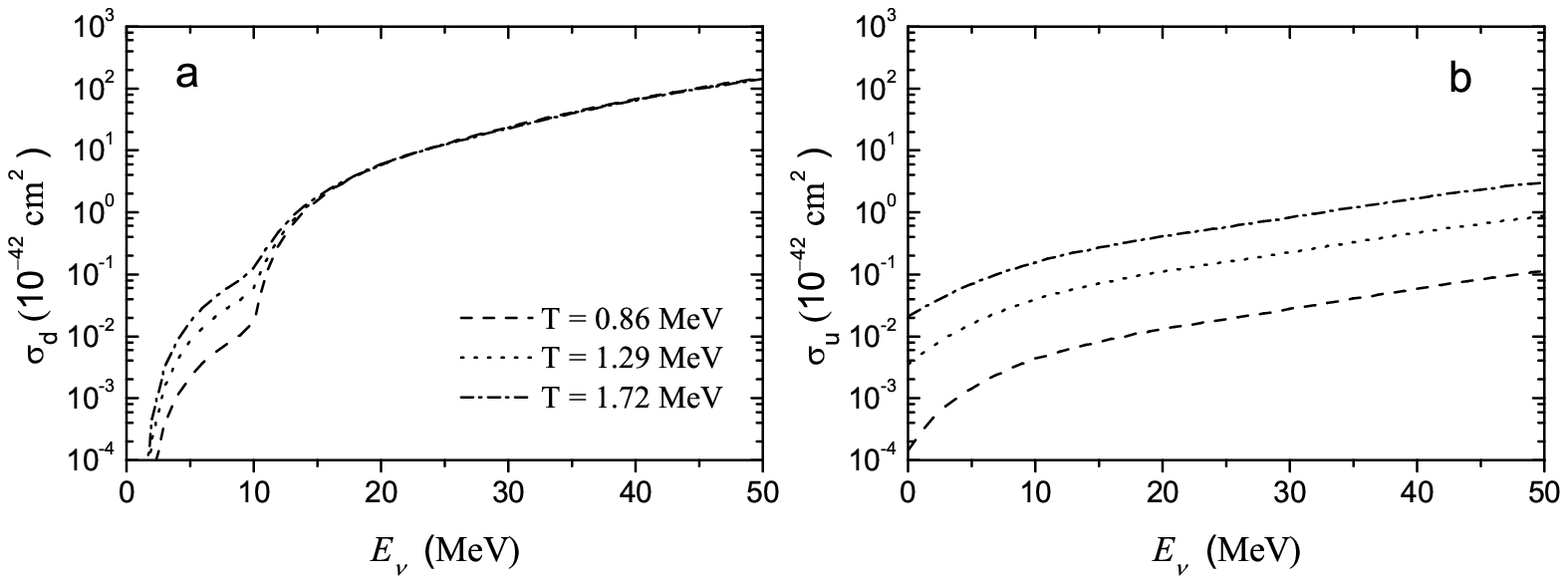}
\caption{}\label{figure5}
\end{figure*}

In the LSSM calculations, the temperature-related enhancement of
$\sigma(E_\nu)$ was only due to the neutrino up-scattering. In our
approach both the up-scattering and down-scattering parts of
$\sigma(E_\nu)$ are temperature dependent. To analyze the relative
importance of these two types of  scattering processes we
display them separately  as the functions of $E_\nu$ for different
values of $T$ in Fig.~\ref{figure5}.

A weak $T$-dependence of $\sigma_d$ is seen at low neutrino
energies $E_\nu < 12$~MeV. At higher energies $\sigma_d$ practically
does not depend on $T$. As the function of $E_\nu$ the
down-scattering cross section sharply increases at low neutrino
energies and then grows more slowly. Instead, $\sigma_u$
is quite sensitive to temperature but its dependence on $E_\nu$ is
obviously smoother than that of $\sigma_d$ (at least at $E_\nu <
15$~MeV). The absolute values of $\sigma_d$ and $\sigma_u$ are of
the same order of magnitude only at quite low neutrino energies
$E_\nu \lesssim 4-10$~MeV.

Thus the conclusion is that the $T$-dependence of the INNS cross
section at low neutrino energies is mainly due to up-scattering
process whereas at neutrino energies $E_\nu > 15$~MeV when the
thermal effects are much less important  the INNS cross section is
determined by the neutrino down-scattering.

The above conclusions agree well with the results of the LSSM
studies for even-even nuclei \cite{PLB02,NPA05}. Furthermore, our
results for $\sigma_d$ confirm the applicability of approximations
based on the Brink hypothesis, which has been used in calculations
of $\sigma_d$ in the LSSM.

\section{Conclusions}

We have performed studies of the temperature dependence of the cross
section for inelastic neutrino-nucleus scattering off the hot nucleus
$^{54}$Fe. Thermal effects were treated within the thermal
quasiparticle random phase approximation in the context of the TFD
formalism. These studies are relevant for supernova simulations.

In contrast to the large-scale shell-model studies
\cite{PLB02,NPA05} we do not assume the Brink hypothesis when
treating the down-scattering component of  the cross section
$\sigma(E_\nu)$. Moreover, we take into account thermal effects
not only for the allowed $1^+$ transitions but also for the
first-forbidden transitions $0^-,~1^-$, and $2^-$. For all
multipole contributions we have performed the calculations with momentum
dependent multipole operators.

Despite these differences between the two approaches, our
calculations  have revealed the same thermal effects as were found
in~\cite{PLB02,NPA05}: A temperature increase leads to a
considerable enhance of the INNS cross section for neutrino energies
lower than the energy of the GT$_0$ resonance. This enhancement is
mainly due to neutrino up-scattering at finite temperature.
The calculated cross sections for $^{54}$Fe are very close
to those given in \cite{NPA05}. Thus, the results of our study show
that the present approach provides a valuable tool for the
evaluation of the inelastic neutrino-nucleus cross sections under
stellar conditions. The approach can be easily adopted to calculate
the INNS cross sections as a function of scattering angle.

\section*{Acknowledgments}

The fruitful discussions with K.\,Langanke and G.\,Mart\'inez-Pinedo
are gratefully acknowledged. This work is supported in part by the
Heisenberg-Landau Program and the DFG grant (SFB 634).


\newpage

\centerline{\bf Figure captions}

\bigskip
\textbf{Fig. 1.} GT$_0$ strength distributions  in the $^{54}$Fe
nucleus at different temperatures $T$ as a function of the energy of
transition $\omega$.

\bigskip
\textbf{Fig. 2.} \textbf{a} -- Contribution of $1^+$ transitions to
the cross section of neutrino  inelastic scattering off $^{54}$Fe
calculated with the $q$-dependent $1^+$ excitation operator as a
function of neutrino energy $E_\nu$  at different stellar
temperatures $T$; \textbf{b} -- A comparison of the cross sections
of neutrino inelastic scattering off $^{54}$Fe calculated with the
$q$-dependent $1^+$ excitation operator (solid line) and the GT$_0$
excitation operator \eqref{GT0} (dashed line) at $T=0.86$~MeV.

\bigskip
\textbf{Fig. 3.} Contributions of different first-forbidden
transitions to the neutrino-nucleus inelastic scattering cross
sections for $^{54}$Fe at different temperatures $T$: \textbf{a} --
the contribution of the $0^-$-transitions; \textbf{b} -- the
contribution of the $1^-$-transitions; \textbf{c} -- the
contribution of $2^-$-transitions; \textbf{d} -- the summed
contribution of the all first-forbidden transitions.

\bigskip
\textbf{Fig. 4.} The neutrino-nucleus inelastic scattering cross
sections as the sum of allowed and first-forbidden contributions for
$^{54}$Fe at different  temperatures $T$.

\bigskip
\textbf{Fig. 5.} The down-scattering $\sigma_d(E_\nu)$ \textbf{(a)}
and the up-scattering $\sigma_u(E_\nu)$ \textbf{(b)} parts of the
neutrino-nucleus inelastic scattering cross section for $^{54}$Fe at
different $T$.

\newpage

\newpage

\newpage

\newpage

\newpage

\end{document}